\documentstyle[preprint,pra,aps]{revtex}
\begin{document}
\preprint{\today}
\draft
%
%
%
\title{Quantum damping of position due to energy measurements}

\author{Roberto Onofrio and Lorenza Viola}
\address{Dipartimento di Fisica ``G. Galilei'' and INFN, 
Universit\`a di Padova,\\
Via Marzolo 8, Padova, Italy 35131}

\date{\today}

\maketitle
%
%
\begin{abstract}
Quantum theory for measurements of energy is introduced and its 
consequences for the average position of monitored dynamical 
systems are analyzed. It turns out that energy measurements 
lead to a localization of the expectation values 
of other observables. This is manifested, in the case of 
position, as a damping of the motion without classical analogue.  
Quantum damping of position for an atom bouncing on a reflecting surface 
in presence of a homogeneous gravitational field is dealt in detail and 
the connection with an experiment already performed 
in the classical regime is studied.
We show that quantum damping is testable provided that the same measurement 
strength obtained in the experimental 
verification of the quantum Zeno effect in atomic spectroscopy 
[W. M. Itano et al., Phys. Rev. A {\bf 41}, 2295 (1990)] is made available.
\end{abstract}
%
%
%
%
\section{Introduction}

A number of new experimental techniques involving devices with noise 
figures close to the minimum dictated by the uncertainty principle 
has originated the demand for quantitative predictions of quantum 
measurement theory \cite{GREEN,BRAGINSKY}.
One of the most important steps along this path was made by Itano 
et al. \cite{ITA}: they experimentally showed that the effect of the 
measurement on the observed system leads to a freezing of its free dynamical 
evolution, the so-called quantum Zeno effect \cite{MIS1,MIS2}. 
In such a case one is looking at the occupancy 
probability of one level, that happens to depend upon the process of 
measurement itself.  
More general quantum measurement effects can be obtained by measuring 
one observable and by looking at the dynamical evolution of another observable, 
as it has been already discussed for a mesoscopic structure in \cite{TAPREON}. 
This agrees with the original spirit with which Sudarshan 
opened the debate on the quantum Zeno effect and 
according to which measurement of the position of a decaying 
particle should influence its lifetime, i.e. another observable quantity.

In this paper we discuss a model for a system subjected to quantum 
measurements of energy including their influence  on other observables, 
specifically position measurements.
The act of measurement affects the average position of the system giving 
rise to an average localization of the motion, a 
{\sl quantum damping} without classical counterpart. 
The paper is organized as follows. 
In Section II we introduce the general formalism for measurements 
of energy distinguishing between nonselective and selective cases 
and we evaluate the average position of the 
measured system in both schemes. The example of the harmonic oscillator, 
for which analytical evaluations are possible, is also dealt. 
In Section III a particle in a homogeneous gravitational field 
bouncing on a perfectly reflecting surface is analyzed in detail.
In particular, the dependence of the quantum localization effect upon the 
relevant parameters is studied both by means of numerical tools as well as in 
the semiclassical limit, by exploiting the WKB approximation. 
In Section IV possible scenarios to look for the predicted quantum damping 
are discussed with particular emphasis on the previously considered particle 
bouncing on a reflecting surface. We focus on a case very close to the 
experimental situation already reported in \cite{AMINOFF}, giving a 
discussion of the parameters that have to be reached to look for 
the predicted effect.

\section{Quantum measurements of energy: general formalism}

Since the original proposal of Von Neumann \cite{VONNEUM}, important 
progresses have been made in understanding quantum measurement theory.
 In particular, it has been recognized that the dynamics of the 
system in presence of the measurement undergoes modifications, with 
respect to the closed system dynamics, that can be taken into account 
by means of an effective master equation or a Schr\"odinger equation for 
mixed and pure states respectively. Thus the original doubling 
of dynamics, a free Schr\"odinger evolution during the non measurement 
periods and an abrupt state collapse during an ideal, instantaneous 
measurement, has been replaced by a unique dynamical approach.
In this context, some models represent 
the meter as a particular environment that interacts 
with the observed system and extracts information from it.
The dynamics of a system interacting with an environment is most conveniently
described in terms of a reduced density matrix operator $\hat \rho(t)$, 
obtained by tracing out the degrees of freedom of the environment 
from the density matrix 
operator of the entire system (marginalization procedure).
The unitary evolution of $\hat \rho(t)$ for the isolated system is 
modified to an irreversible one due to the interaction with the environment. 
In the limit of a Markovian environment, a dynamical law described by a
completely positive semigroup can be postulated 
for the evolution of the open system \cite{LINDBLAD,GORINI},
resulting in the following
master equation for the reduced density matrix operator:
\begin{equation}
{d \over dt} \, \hat\rho(t) = 
- {i \over \hbar} \left[ \hat H(t), \hat \rho(t) \right] 
+ {1\over2} \sum_{\nu=1}^n \left( 
\left[ \hat L_\nu(t) \hat \rho(t) , \hat L_\nu(t)^{\dag} \right] +
\left[ \hat L_\nu(t) , \hat \rho(t) \hat L_\nu(t)^{\dag} \right]
\right) 
\label{LINDBEQ}
\end{equation}
where $\hat H(t) = \hat H(\hat p, \hat q,t)$ is the Hamilton operator
for a general nonautonomous system and $\hat L_\nu(t),\:\nu=1,\ldots,n$, are 
the so-called {\sl Lindblad operators}, that are supposed to model the effects 
of the environment on the system.
The above equation describes the general case of a quantum open system. 
The evolution of a quantum system subjected to a measurement process 
corresponds to the particular case where the environment is the measurement 
apparatus and the Lindblad operators are proportional to the measured 
quantities.
If the measurement of a single observable 
represented by the operator $\hat A(t) = \hat A(\hat p, \hat q,t)$
is considered, the corresponding Hermitian Lindblad  
operator can be chosen to be 
$\hat L(t) \equiv \hat L(t)^{\dag} = \sqrt{\kappa(t)} \hat A(t)$. 
The function $\kappa(t)$ has dimensions $[\kappa]=[t^{-1} A^{-2}]$ 
and represents the coupling, in general time-dependent,  
of the monitored system to the measurement apparatus.
The result of the measurement is 
\begin{equation}
\overline{a(t)} = {\rm Tr} \left( \hat{A}(t) \hat\rho(t) \right)\;, 
\label{AVENS}
\end{equation}
where overlining denotes a statistical average over individual results 
associated to pure states that form the incoherent mixture
described by the reduced density matrix operator $\hat\rho(t)$. 

By moving to the coordinate representation of the reduced density matrix 
operator, 
\begin{equation}
\rho(q_1,q_2,t) =  \langle q_1 |~\hat \rho(t)~|q_2 \rangle\;, 
\end{equation}
the evolution equation obtained from (\ref{LINDBEQ}) is 
\begin{eqnarray}
{\partial \over \partial t} \, \rho(q_1,q_2,t) = 
\Biggl\{ &-&         
{i \over \hbar} H\left( -i \hbar {\partial \over \partial q_1},q_1,t \right) 
+{i \over \hbar} H\left( -i \hbar {\partial \over \partial q_2},q_2,t \right)   
\nonumber \\ 
&-& {1 \over 2} \kappa(t) 
\left[ A\left( -i \hbar {\partial \over \partial q_1},q_1,t \right) - 
       A\left( -i \hbar {\partial \over \partial q_2},q_2,t \right) \right]^2
\Biggr\} \rho(q_1,q_2,t) \;. \label{EVOL} 
\end{eqnarray}

Equation (\ref{EVOL}) gives the general description of a 
system in which the observable $\hat{A}$ is continuously monitored and 
the result of the measurement is not known in advance, the so-called {\sl 
nonselective} measurement process. 

If continuous measurements of energy characterized by a constant coupling 
$\kappa_{E}$ are considered ($\hat{A}=\hat{H}$), 
Eq. (4) for the evolution of the density matrix specializes to 
\begin{equation}
\dot{\rho}_{nm}=-{i\over \hbar} (E_n-E_m)\rho_{nm} - {\kappa_{E} \over 
2} 
(E_n-E_m)^2 \rho_{nm}\;,
\label{rho}
\end{equation}
where 
\begin{equation}
\rho_{nm}(t)=\int \int~ dq~dq'~\rho(q,q',t)\phi_n^*(q) \phi_m(q')
\label{rho1}
\end{equation}
and $\{ \phi_n\}$ is the complete set of energy eigenstates, 
corresponding to the energy eigenvalues $E_n$. 
It is easy to show that the solution of the equation (\ref{rho})
can be written as 
\begin{equation}
\rho_{nm}(t)=\exp\left\{{-{i \over \hbar} (E_n-E_m)t -{\kappa_E 
\over 2} (E_n-E_m)^2 t}\right\}
\rho_{nm}(0)\;,
\label{DECOH}
\end{equation}
and consequently 
\begin{equation}
\rho(q,q',t)=\sum_{nm} \rho_{nm}(t) \phi_m^*(q')\phi_n(q)\;.
\label{DENSITY}
\end{equation}
Notice that the effect of the measurement in (\ref{DECOH}) 
is to diagonalize the density matrix after enough time,
the so-called {\sl decoherence} induced by the measurement \cite{ZUREK}.

Without loss of generality, we can restrict our analysis to the position 
operator $\hat{Q}$. 
The average position at time $t$ is then evaluated as
\begin{equation}
\langle Q(t) \rangle =\int dq~dq'~\rho(q,q',t)~\langle q'|\hat{Q}|q \rangle 
=\int dq~\rho(q,q,t)~q
\end{equation}
and therefore from (\ref{DENSITY})
\begin{equation}
\langle Q(t) \rangle=\sum_{nm} \rho_{nm}(t) \langle n |\hat{Q} |m\rangle 
\;.  \end{equation}
By inserting (\ref{DECOH}),  
the average position in the nonselective case we have just considered is 
finally written as
\begin{equation}
\langle Q(t) \rangle=\sum_{nm} 
\exp\left\{{-{i\over \hbar} (E_n-E_m)t-{\kappa_E \over 2} 
(E_n-E_m)^2t}\right\}\rho_{nm}(0)\langle n |\hat{Q}|m\rangle \;,
\label{NSEL}
\end{equation}
where the effect of the measurement leads to exponential decaying 
behaviour with state-dependent time constants
\begin{equation}
\tau_{nm}={2 \over {\kappa_{E}(E_n-E_m)^2}}.
\end{equation} 
It is worth to observe that the off-diagonal terms of the sum contributing to 
the average position  
vanish in the asymptotic limit $t\rightarrow \infty$ 
when $\kappa_E>0$. 
As a consequence, the average position tends to be localized around the 
stationary value due to the diagonal contributions.
From the conceptual viewpoint, this effect is nothing but the 
manifestation, in the configuration space, of the previously mentioned 
decoherentization process.

Another interesting situation arises when the energy measurement 
is already performed and the result is known. The theory should be able, in 
this case, to complete the knowledge of the system by evaluating 
the corresponding wavefunction conditioned to the known result 
of the energy measurement, the so-called 
{\sl a posteriori selective} measurement.
In \cite{PREONTA} it is shown that, starting from (\ref{EVOL}), an 
effective Schr\"odinger equation for an {\sl a posteriori} measurement 
is obtained for the restricted wavefunction $\psi_{[a]}(q,t)$ as:
\begin{equation}
i \hbar \,{ \partial {\psi_{[a]}(q,t)} \over {\partial t}}=
\left[
 H(-i\hbar {\partial \over \partial q}, q, t)
-i \hbar\, \kappa_{[a]}(t) 
\left[ A(-i\hbar {\partial \over \partial q},q,t)-a(t) \right]^2 \right] 
\psi_{[a]}(q,t) \label{PSISEL}
\;,\end{equation} 
where we have denoted by $a$ the result of the measurement and by $[a]$ 
the associated functional dependence. 
Equation (\ref{PSISEL}) formally holds also when the result 
$a(t)$ is unknown and has to be 
predicted with probability distribution $\|\psi_{[a]}(t)\|^2$ 
({\sl a priori selective measurement}); a different interpretation of 
(\ref{PSISEL}) is however required, for which we refer to \cite{PREONTA}. 
Coming back to continuous measurements of energy with a constant 
result $E(t)=E$, Eq. (\ref{PSISEL}) has the solution \cite{TAPREON,ONPRETA}
\begin{equation}
\psi_{[E]}(q,t)=
\sum_n c_n(0)\exp\left\{{-iE_n 
t/\hbar-\kappa_E(E_n-E)^2t}\right\}\,\phi_n(q) \;.
\label{SELEC}
\end{equation}
Here, as before, $\phi_n(q)$ are energy eigenfunctions, $c_n(0)$ the 
corresponding initial projection coefficients, and 
normalization to unity is lost due to the non Hermitian character 
of Eq. (\ref{PSISEL}), that expresses in more physical terms the 
branching of the wavefunction among all the possible alternatives. 
By exploiting (\ref{SELEC}), the average position in 
this selective case, hereafter 
denoted by $\langle Q(t) \rangle_{[E]}$, is therefore written as 
\begin{eqnarray}
\langle & & \hspace{-2mm}Q(t) \rangle_{[E]} = 
\langle \psi_{[E]} | \hat{Q} | \psi_{[E]} \rangle /
\langle \psi_{[E]}| \psi_{[E]} \rangle= 
{ \biggl[ \sum_k |c_k(0)|^2 
\exp\{ -2\kappa_E(E_k-E)^2\,t \} \biggr] }^{-1} \cdot  \nonumber \\
& \cdot & \hspace{2mm}\sum_{nm} c_n(0) c_m^*(0) 
\exp\left\{ -i(E_n-E_m)t/\hbar- \kappa_E[(E_n-E)^2+(E_m-E)^2]\,t  \right\}  
\, \langle m|\hat{Q}|n \rangle \;. \label{SEL}
\end{eqnarray}
The link between the average position in the nonselective (\ref{NSEL}) 
and selective (\ref{SEL}) cases is established, in the usual way 
\cite{PREONTA}, by summing over all the possible selective measurement 
processes, i.e. 
\begin{equation}
\rho(q,q',t)=\int~d[E]~\psi^*_{[E]}(q,t)~\psi_{[E]}(q',t)\;,
\end{equation}
that also shows how the nonselective measurement, described in terms of 
the density matrix, can be seen as a functional integration over all the 
possible selective measurements, represented by the 
restricted wavefunction. Note also that the decay time constants in 
(\ref{SEL}) explicitly depend upon the value $E$ registered during the 
measurement.

The theoretical scheme we have presented can be easily implemented for a 
simple system as the harmonic oscillator. Let us denote by 
$m$ and $\omega$ the mass and the angular frequency respectively. The effect 
of the measurement on the average position in the nonselective case can be 
guessed by simple inspection of Eq. (\ref{NSEL}). Indeed, due to the 
equal spacing between adiacent levels, one immediately realizes that the 
measurement damping effect factors out and acts on the remaining unmeasured 
evolution as a purely exponential damping with an unique time 
constant $\tau=2/\kappa_E \hbar^2 \omega^2$, that is: 
\begin{equation}
\langle Q(t) \rangle_{\kappa_E}= 
\exp\left\{ -{{\kappa_E (\hbar \omega)^2 t} \over {2}} \right\}\;
\langle Q(t) \rangle_{\kappa_E=0}\;. \label{FACTORIZATION}
\end{equation}

In the selective case instead the corresponding expression (\ref{SELEC}) 
can be explicitly computed if, as usual, creation and annihilation 
operators $\hat{a}, \hat{a}^{\dagger}$ are introduced so that
\begin{equation}
\hat{Q}=\sqrt{\hbar\over{2m\omega}}(\hat{a}+\hat{a}^{\dagger})\;.
\end{equation}
\noindent Then, owing to the well known selection rule for the matrix 
elements of the position operator between energy eigenstates, namely
\begin{equation}
\langle n|\hat{Q}|m \rangle=\sqrt{\hbar\over{2m\omega}}
\: \langle n|(\hat{a}+\hat{a}^{\dagger})|m \rangle=
\sqrt{\hbar\over{2m\omega}}\: (\sqrt{m}\,\delta_{n,m-1}+
\sqrt{m+1}\,\delta_{n,m+1})\;,
\end{equation}
\noindent the following expression for the average position is found:
\begin{eqnarray}
& & \langle Q(t) \rangle_{[E]}  =  2 \left({\hbar \over {2m\omega}}\right)^{1/2}
{\biggl\{\sum_k |c_k(0)|^2 \exp [-2 \kappa_E (E_k-E)^2 t]\biggr\}}^{-1} 
\cdot \sum_{n=0}^{\infty} 
\sqrt{n+1} |c_n||c_{n+1}| \cdot \\ \nonumber 
 & \cdot & \exp \left\{ -2\kappa_E \left[ 
\left[ \left( n+\frac{1}{2} \right) \hbar \omega-E \right]^2+
\hbar \omega \left[ \left(n+\frac{1}{2} \right) \hbar \omega-E \right]+
\frac{\hbar^2 \omega^2}{2} \right] t \right\}
 \cos(\omega t+\theta)\;,
\label{SELEQUA}
\end{eqnarray} 
\noindent where $\theta$ is the relative phase between $c_n$ and $c_{n+1}$.
Unlike the nonselective case, no factorization of 
the damping factor is allowed here. 
By referring to the decay constant for the nonselective measurement 
$\tau$, we get
\begin{equation}
\tau_{[E]}(n)={\tau \over 4} 
\left[ \left(n+{1 \over 2}-{E \over {\hbar \omega}} \right)^2+
n+1-{E \over {\hbar \omega}} \right]^{-1}\;.
\end{equation}
The time constants will now depend upon the registered 
energy $E$, decreasing with the difference between the ratio 
$E/\hbar \omega$ and the average number of quanta in the state.

\section{Free fall of a particle and bouncing on a reflecting surface}

Another stimulating example is the monitoring of the energy of a particle 
falling in an homogeneous gravitational field and bouncing 
on an elastically reflecting surface. Experiments already performed 
have shown that multiple bouncing of atoms on surfaces is possible 
\cite{AMINOFF} and the classical sources of damping have been 
understood \cite{BALYKIN}. 
The experiment has been performed by measuring the average of an ensemble 
of independent atoms and therefore the nonselective approach described in 
Section II is more adequate.

Let us consider the 
height  $z$ over the surface and the potential energy
\begin{equation}
V(z)= \left\{ \begin{array}{cl}
m\,g\,z & \hspace{2mm}z>0 \;, \\
+\infty & \hspace{2mm}z \leq 0 \;.  \end{array} \right.
\end{equation}
The Schr\"odinger equation for positive $z$ and energy eigenvalue $E$ is 
\begin{equation}
-{\hbar^2\over {2m}}{d^2 \psi \over dz^2}+(mgz-E)\psi=0
\label{AIRYEQ}
\end{equation}
and we assume, consistently with the boundary conditions of the problem, 
$\psi(z)=0$ for $z \leq 0$. Equation (\ref{AIRYEQ}) is exactly solvable and 
the solutions are expressed in terms of the Airy functions \cite{FLUGGE}
\begin{equation}
\phi_n(z)=C_n~Ai\,(z/z_0-\lambda_n)\;,
\end{equation}
where we have introduced the characteristic length of the system 
\begin{equation}
z_0 ={ \left(\hbar^2\over {2 m^2 g}\right) }^{1/3}\;.
\label{length}  \end{equation}
The $C_n$'s are normalization constants and $-\lambda_n$ 
is the $n$-th zero of the Airy function, related 
to the corresponding energy eigenvalue through the following relation:
\begin{equation}
E_n={\hbar^2\over {2m z_0^2}}\,\lambda_n \;.
\end{equation}
According to these conventions, $z_0\lambda_n$ represents the $n$-th 
classical turning point, also denoted by $z_n$.

The distance between consecutive zeroes of the Airy functions 
$|\lambda_n-\lambda_{n+1}|\rightarrow 0$ for $n\rightarrow \infty$. 
It is therefore recognizable that, for this system, unlike the harmonic 
oscillator, the quantum damping is large for states 
formed by low-energy pairs of consecutive eigenstates and vanishes in 
the (classical) limit of high energy. 
To understand the mechanism of damping we have numerically  studied 
various cases corresponding to different initial preparations of the system.
The numerical accuracy of the program, tested with the already analytically 
solved harmonic oscillator, is of the order of $0.1\%$ for reasonable values 
of the space-time lattice.
If energy eigenstates are considered, the average position is constant in 
time but, unlike the case of the harmonic oscillator, it is different from 
zero due to the nonvanishing diagonal matrix elements of the position 
operator. However, these stationary states are not 
affected by the presence of a measurement coupling. On the other hand, 
when a superposition of two energy eigenstates is assigned at zero time, 
the average position will 
harmonically oscillates  between a minimum and a maximum value with 
angular frequency $\omega=(E_n-E_m)/\hbar$ ($n=2, m=1$ in the example 
of Fig. 1a). 
By comparison, the constant values of the average position 
in the two energy eigenstates are also shown, the quantities 
\begin{equation}
Q_{11}=\int dz \, \phi_1(z)^*\,z\,\phi_1(z)~~;~~
Q_{22}=\int dz \, \phi_2(z)^*\,z\,\phi_2(z) \;,
\end{equation}
which allows one to write the average position of their superposition 
$\psi=c_1\phi_1+c_2\phi_2,$ as 
\begin{equation}
\langle Q(t) \rangle=|c_1|^2~Q_{11}+|c_2|^2~Q_{22}
+2|c_1||c_2|Q_{12}\cos\left[{{(E_1-E_2)t}\over{\hbar}}+\theta\right] \;,
\label{SUPERP}
\end{equation}
with
\begin{equation}
Q_{12}=\int dz \, \phi_2(z)^*~z~\phi_1(z)=Q_{21}^*\;.
\end{equation}
In presence of a measurement a pure exponential damping arises that 
still reminds the oscillatory behaviour (as shown 
in Fig. 1b) or that is overdamped (as in Fig. 1c), 
depending upon the measurement coupling through the 
time constant $\tau_{12}=2\,[{\kappa_E(E_2-E_1)^2}]^{-1}$. The time 
development is centered around the mean value of the unmeasured 
evolution (\ref{SUPERP}). The effect of the energy measurement vanishes 
in the limit of large quantum numbers, namely the classical limit. This 
can be easily shown by exploiting the WKB approximation, that gives the 
spacing between two consecutive levels as reported in \cite{WALLIS}
\begin{equation}
E_n-E_{n-1}=\pi \hbar \left( {g \over {2z_n}} \right)^{1/2}\;.
\end{equation}

Since it is difficult to obtain superpositions of only two 
energy eigenstates, the attention must be focused on states that are more 
likely to be produced when an atomic 
cloud is prepared. This is the case of the Gaussian states, 
already extensively studied for being the quantum states 
closest to classicality, although 
still far from representing an atomic cloud. 
As discussed in \cite{FOOT}, 
Cesium atoms in magnetooptical traps lead at best to an initial  
radius of the cloud equal to $z_i\simeq 50$ $\mu$m with an 
rms momentum corresponding to a velocity spread of 
$\simeq 2$\,cm/s, leading to an uncertainty product 
$z_i p_{z_i} \simeq 2000 \,\hbar$. Nevertheless, Gaussian states constitute 
entities simple enough to be analyzed, allowing a physical understanding 
of their behaviour, and at the same time complex enough to maintain 
the typical features of the more realistic situations. 
Indeed, the striking difference from the already analyzed case is that 
now the energy eigenstates expansion cointains different eigenstates 
and a more complex dynamics is obtained also in the 
unmeasured case (Fig. 2a). When the effect of the 
measurement is taken into account, a damping of the motion is obtained 
in all the explored cases, although it is not a pure exponential 
damping, as shown from the weak reformation of the damped oscillations 
in Fig. 2b. By varying the height of the atomic center of mass 
for a Gaussian state the relevant contributing eigenstates also change.
In general, as also intuitively understandable, for a height 
$z$ the greatest contributions will stem from the eigenstates having 
the classical turning point $z_0 \lambda_n$ closest to $z$.
Moreover, the number of contributing eigenstates will depend upon the 
width of the initial Gaussian state.
In Fig. 3 a configuration similar to Fig. 2 but with a larger 
width of the Gaussian state is depicted. In this case, more energy 
eigenstates significantly contribute to the expansion and in particular the 
lower ones determine, due to their larger energy difference, a 
faster dynamics for the average position. As a general feature, among the 
various time constants contributing to the damping, the more relevant 
ones correspond to the eigenstates that at the same time possess larger 
energy separation and appreciable contribution to the state itself. 

Also here the WKB approximation allows one to estimate the damping time 
constants. By considering a rough 
picture in which a state is only made of eigenstates centered around 
$\bar{n}$ with width $2\,\Delta \bar{n}$, all contributing with the 
same weight, the dynamics is ruled by the farest energy eigenvalues:  
\begin{eqnarray}
\tau_{min}& = & {2 \over {\kappa_E  {|E_{\bar{n}+\Delta\bar{n}}-
E_{\bar{n}-\Delta \bar{n}}| }^2 }}  \nonumber \\ 
          & = & { \biggl( \frac{2}{3 \pi}\biggr) }^{4/3}
  \frac{8 m^2 {z_0}^4}{\kappa_E \hbar^4} 
{ \biggl[ { \biggl( \bar{n} + \Delta \bar{n} - {1 \over 4} \biggr) 
           }^{2/3}  - 
{ \biggl( \bar{n} - \Delta \bar{n} - {1 \over 4} \biggr) 
           }^{2/3} \biggr] }^{-2}  \;.  
\label{TAUNONSEL}
\end{eqnarray}

\section{Phenomenological considerations}

In this section we analyze observable consequence of the measurement of 
energy in the two cases of harmonic oscillators and bouncing particles 
in the presence of gravitational fields.
Before doing this, some preliminar considerations are 
needed about the relevant coupling constant of the model, $\kappa_E$.
Bearing in mind that this parameter, describing the effective coupling of 
the apparatus to the observed system, intrinsically depends upon the 
particular experimental setup, one can only assess for it a reasonable value. 
If the analysis for a continuous nonselective measurement of energy is 
applied to the results of Itano et al. as done in \cite{PREONTA}, 
a lower bound for the coupling parameter $\kappa_E$ is found. 
In \cite{PREONTA} the coupling parameter is expressed in terms of a 
critical value $\kappa_{crit}=4 \omega_R$, where $\omega_R=12.272 s^{-1}$ 
is the Rabi angular frequency (corresponding to a period of the 
radiofrequency of $256\, ms$) 
and the energy difference between the two levels $E_2-E_1=2.125 \cdot
10^{-25}J$. The data of the experiment are fitted for $\kappa_E^{exp} 
\approx 10^2 \kappa_{crit} / (E_2-E_1)^2 = 
10^{53}\,J^{-2}s^{-1}\;$. 
This value will be assumed, in some of the following examples, 
as an indicative one.

  It is worth to note that, due to the generality of the model, observable 
effects are in principle expected in all physical situations where 
quantum measurements are involved. Among all the 
possible systems in which the effect can be made observable, we have 
chosen to discuss atomic or
molecular systems and single degrees of freedom of macroscopic bodies. 

Let us first reconsider the case of systems modelizable as harmonic 
oscillators. 
If, for example, the vibrational energy levels of a biatomic 
molecule are monitored and the position operator $\hat{Q}(t)$ 
is interpreted as describing the istantaneous electric dipole 
moment of the molecule along the internuclear 
axis, then, according to (\ref{FACTORIZATION}), one can roughly expects a 
modification of the law according to which emission and absorption of 
electromagnetic radiation occurs. For instance, the intensity $I(t)$ 
will not decay with an exponential law ruled by a lifetime $\tau$, but 
it will manifest a more complex behaviour:
\begin{equation}
I(t)=I_0 \exp\Big\{-{\exp\big[ \kappa_E {(\hbar \omega)}^2 t\big]{t \over \tau}}
\Big\} 
\end{equation} 
corresponding to an inhibition of the decay.
Unfortunately, the time scale resulting when the 
$\kappa_E^{exp}$ quoted before is of the order of $10^8 s$ for 
reasonable vibrational frequencies, prohibitively long compared 
to the typical lifetimes of the vibrational transitions.
Other perspectives can be opened by considering the time and frequency 
resolved spectra of spontaneous emission recently demonstrated 
\cite{DUNN}. 

Alternatively one could measure the energy in mechanical harmonic oscillators 
such as the resonators used as gravitational wave antennas, provided that the 
quantum limit is achieved in such a class of detectors \cite{BOCKO,MAJORANA}. 
Due to the presence of a single system the analysis should be carried out 
by using the selective measurements approach.
In this case, however, one big problem is the difficulty 
to obtain electromechanical transducers that measure the energy 
of the oscillator with enough sensitivity \cite{CAVES}.

 Other possibilities are also open by exploiting the quantum measurement 
model applied to the bouncing particle of Section III. 
An experiment aimed at testing the 
quantum damping can be designed on the basis of already performed experiments 
as the ones described in \cite{ITA} and in \cite{AMINOFF}. A cloud of atoms 
is trapped and cooled at a given height over a dielectric surface as in 
\cite{AMINOFF}. Along the vertical path we put both a inhomogeneous 
magnetic field with constant gradient and a radiofrequency. 
The magnetic field is such that at the initial height the radiofrequency 
gives rise to resonant Rabi oscillations between two levels of the atoms. 
The hyperfine splitting changes with the magnetic field and for each 
height less than the initial one the resonant condition is not fulfilled.
A continuously operating laser acting along the vertical direction 
is tuned to the optical transition between level 1 and a third level 
which has forbidden transitions with level 2. A set of optical detectors 
allows one to observe the fluorescence light proportional to the occupancy 
of level 1. On the other hand, as 
shown in \cite{PREONTA}, the occupancy of a level can be also thought as 
a measurement of energy since the state projectors of the occupancy and 
the energy operators coincide apart from a dimensional constant. Due to 
the spatially variable magnetic field, the continuous measurement of the 
occupancy will push the state of the atoms toward gravitational energy 
eigenstates and this will affect their average position. 
The damping of the average position of the atomic cloud could be measured 
with a destructive probe photon beam by repeating the measurements 
many times, as in \cite{AMINOFF}, or by exploiting non-destructive 
measurement schemes, such as the quantum nondemolition dispersive 
atomic probe one \cite{BRUNE}, allowing to repeatedly monitor the same 
atomic cloud (for more recent proposal see also \cite{COURTOIS,ASPECT}). 
An estimate of the quantum damping time can be given by using 
Eq. (\ref{TAUNONSEL}) once expressed in terms of the more accessible 
position variance of the atomic cloud. 
The energy spread due to the $z$-motion of the atomic cloud can be written as 
\begin{equation}
\Delta E^2=\langle H^2 \rangle - \langle H \rangle^2 \;,
\end{equation}
where $H=p_z^2/2m +V(z)$. By supposing a Gaussian initial atomic phase space 
distribution both in coordinate and momentum, centered on the 
$z$-axis at an height $z_E$ above the reflecting surface, we have a Wigner 
function
\begin{equation}
W(z,p_z)={1 \over {2\pi z_i p_{z_i}}}\exp{[-(z-z_E)^2/2z_i^2]}\cdot 
\exp[-p_z^2/2p_{z_i}^2]
\end{equation}
and the evaluation of the energy spread gives in this case \cite{WALLIS}
\begin{equation}
\Delta E=m g z_0 \Big[2N^4 \Big({z_0 \over z_i}\Big)^4+
\Big({z_i \over z_0}\Big)^2 \Big]^{1/2}
\end{equation}
where the uncertainty product of coordinate and momentum in 
units of $\hbar$ has been introduced $z_i p_{z_i} / \hbar= N$. 
The energy spread has a value equal to $\Delta E \approx mgz_i$ for large 
values of $z_i/z_0$, as one expects from the classical behaviour. 
In the opposite situation the effect of the Heisenberg principle appears 
and gives an inverse law dependence of the energy width upon the 
initial position spreading and a branching of the curves 
for different values of the  uncertainty product $N$. A minimum 
value of the energy width is obtained for  an intermediate value equal 
to $z_i=2^{1/3} N^{2/3} z_0$.

The quantum decay constant is expressed as 
\begin{equation}
\tau={2 \over \kappa_E \Delta E^2}= {2 \over {
\kappa_E m^2 g^2 z_0^2}} \Big[2 N^4 \Big({z_0 \over z_i}\Big)^4+
\Big({z_i \over z_0}\Big)^2\Big]^{-1}
\end{equation}
and in correspondence to the minimum of the energy spread gets the 
maximum value equal to 
$\tau_{\mbox{max}}(N)=2^{-5/3} \kappa_E^{-1} m^{-2}g^{-2} z_0^{-2} 
N^{-4/3}$.

In Fig. 4 the decay constant versus the ratio $z_i/z_0$ is shown for 
various values of the normalized uncertainty product $N$, including the 
case of a pure state $(N=1/2)$. 
Smaller decay times are observed  either for 
large values of the initial position uncertainty, in the right part of 
all the curves, or in the branching of the curves for the left part, 
in this last case depending upon the normalized uncertainty product. 
In the estimate we have assumed a constant  
width of the atomic cloud, an hypothesis that is not strictly valid 
due to the spreading following its preparation.
To  minimize this spreading a very low starting temperature 
of the atomic cloud is required, 
for instance ${}^{85}Rb$ clouds were shown to double their diameter 
in $15\,$ms if cooled at a temperature of about $10\mu$K \cite{KUNZE}.
Since low temperatures are also associated to small initial diameters 
of the cloud we expect the approximation of constant diameter to better 
hold in the left part of the curves drawn in Fig. 4.
In these cases values of the quantum damping comparable to the 
ones of the right part are obtained for widths 
around ten-one hundred times the fundamental length $z_0$, i.e. 
around $2 \div 20 \mu$m. 
For comparison, the point corresponding to 
the decay observed in \cite{AMINOFF} is reported. 
To attribute the damping to the predicted quantum effect the classical 
sources of damping should be kept as small as possible, due also to 
the dynamics imposed by the Rabi transition frequency. 
Currently achieved values of the escape time of atomic clouds are in the  
range of 10 seconds \cite{RAAB,DAVIDSON} 
and if this figure can be maintained together with a 
measurement coupling of the order of magnitude of the one corresponding 
to the experiment described in \cite{ITA} and 
a normalized uncertainty product $\approx 2\cdot10^5$, quantum 
damping is made observable. 

\section{Conclusions}

A quantum damping without classical counterpart has been introduced 
as a consequence of quantum measurements of energy and discussed 
for two situations. The model of the harmonic oscillator 
could be implemented in the monitoring of the vibrational motion 
of biatomic molecules or in quantum limited measurements in 
macroscopic mechanical resonators. 
Quantum localization can also be manifested using
a cloud of atoms bouncing over a reflecting surface 
in the presence of a uniform gravitational field.
A possible experimental scheme based on this last configuration
has been  discussed in more detail leading to a proposal 
that should merge two experiments already 
separately performed in \cite{ITA} and \cite{AMINOFF}. 
Studies of the quantized structure of particles in gravitational 
fields and the observation of a damping purely connected to 
the effect of the measurement as described here 
give further motivations to improve the cooling capabilities of atomic traps.

%
%

%
%

\begin{figure}
\protect
\caption{Average position versus time in the case of a particle 
bouncing on a reflecting surface for an initial pure state 
superposition of the first two energy eigenstates with 
amplitude coefficients $\protect 1/2$ and $\protect \sqrt{3}/2$ respectively.
The case a) is relative to an unmeasured system ($\kappa_E=0$) 
and the dashed curves represent the constant values for the 
two eigenstates $\phi_1$ (below) and $\phi_2$ (above). 
The cases b) and c) are relative to $\kappa_E=10^\protect{-2}$ and 
$\kappa_E=10^\protect{-1}$ representing two examples in the 
underdamped and overdamped regimes, respectively. Here and in the analogous 
cases of Figs.2 and 3 we put $\hbar=m=1$.}
\end{figure}

\begin{figure}
\protect
\caption{Average position versus time for a particle bouncing 
on a wall and schematized by a Gaussian state whose center is initially 
located at $h=10$ with variance $\sigma=1$. 
In the energy eigenstates expansion the main contributions stem 
from the eigenvalues between the fifth and the nineth. 
In a) the unmeasured case is depicted, in b) a measured case with 
$\kappa_E=10^{-2}$ is shown. Note the persistence of the oscillations 
which increase after a minimum indicating a not pure exponential damping
unlike the one of Fig. 1. The lack of complete periodicity of the 
unmeasured case here and in the following 
Fig. 3a is attributable to the presence of 
many eigenstates contributing to the wavefunction reconstruction on 
timescales longer than the one depicted.}
\end{figure}

\begin{figure}
\protect
\caption{The same as in Fig. 2 but for a Gaussian state with variance 
$\sigma=3$. The larger spreading corresponds to an increase of the 
number of eigenstates which significantly contribute to the state, 
in this case between $n=3$ and $n=12$. Case a) is the unmeasured case, 
case b) is relative to a measurement with $\kappa_E=10^{-2}$.}
\end{figure}
\samepage
\begin{figure}
\caption{Decay constant for the quantum damping versus the position 
spreading corresponding to a Gaussian Wigner distribution of Cesium 
atoms bouncing in a gravitational cavity, 
normalized to the fundamental gravitational length $z_0$ 
(mass $m_{Cs}=2 \cdot 10^{-25}$Kg, gravitational length $z_0=0.23 \mu$m). 
The curves are obtained for different values of the 
normalized uncertainty product of the atomic cloud, respectively 
$N=1/2$ (pure state, a), $N=20$ (b), $N=2\cdot 10^3$ (as 
experimentally achieved in \protect\cite{FOOT}, c), and $N=2 \cdot 10^5$ (d).
For comparison the experimental point 
explored in \protect\cite{AMINOFF}, corresponding to a decay 
time of $\simeq 80 \,$ms for a value of $z_i/z_0 \approx 10^3$ and 
explained in terms of classical sources of damping, is shown. 
It has been assumed a measurement coupling 
constant $\kappa_E=10^{53} J^{-2}s^{-1}$.}
\end{figure}

\end{document}